\documentclass[aps,11pt,onecolumn,prd,preprintnumbers,amsmath,amssymb,nofootinbib,superscriptaddress,showpacs]{revtex4-1}
\pdfoutput=1
\usepackage[T1]{fontenc}
\usepackage{amsthm}
\usepackage{amsfonts}
\usepackage{xcolor}
\usepackage{slashed}
\usepackage{mathtools}
\usepackage{graphicx}
\usepackage{hyperref}
\hypersetup{
    colorlinks,
    citecolor=red,
    filecolor=black,
    linkcolor=blue,
    urlcolor=black
}

\begin{document}

\title{Free energy dependence on spatial geometry for (2+1)-dimensional QFTs}

\author{Krai Cheamsawat}
\email{krai.cheamsawat15@imperial.ac.uk}

\author{Lucas Wallis}
\email{l.wallis17@imperial.ac.uk}

\author{Toby Wiseman}
\email{t.wiseman@imperial.ac.uk}

\affiliation{Theoretical Physics Group, Blackett Laboratory, Imperial College, London SW7 2AZ, United Kingdom}


\begin{abstract}

We consider (2+1)-QFT at finite temperature on a product of time with a static spatial geometry. 
The suitably defined difference of thermal vacuum free energy for the QFT on a deformation of flat space from its value on flat space  is a UV finite quantity, and for reasonable fall-off conditions on the deformation is IR finite too.
For perturbations of flat space we show this free energy difference goes quadratically with perturbation amplitude and may be computed from the linear response of the stress tensor.
As an illustration we compute it for a holographic CFT finding that at any temperature, and for any perturbation, the free energy decreases.
Similar behaviour was previously found for free scalars and fermions, and for unitary CFTs at zero temperature,  suggesting  (2+1)-QFT may generally energetically favour a crumpled spatial geometry.
We also treat the deformation in a hydrostatic small curvature expansion relative to the thermal scale. Then the free energy variation is determined by a curvature correction to the stress tensor and for these theories is negative for small curvature deformations of flat space.

\end{abstract} 

\maketitle


\section{Introduction}

The vacuum energy of a relativistic quantum field theory on a static spacetime provides an energy measure on the geometry. We might then ask what type of geometry a QFT prefers energetically. An important subtlety is that the one-point function of the stress tensor must be renormalised, and this 
introduces ambiguity
 into the resulting energy. Typically the leading UV divergence in this vacuum expectation value is removed and the ambiguity in the counterterm, a cosmological constant, is chosen so that the one point function vanishes on flat spacetime. However, there are subleading divergences which require local curvature counterterms to remove, and their finite remainder can't be fixed by simply requiring vanishing on flat spacetime since these counterterms  trivially vanish there. Thus, a notion of energy, or equivalently at finite temperature, free energy, in curved spacetime is ambiguous unless one has a UV complete theory.
Worse still, even on flat space if the theory is renormalised to have vanishing energy at zero temperature, then at finite temperature the total free energy will be IR divergent due to infinite volume, leading one typically to work instead with free energy density.

While the free energy is ambiguous, and on a non-compact space will generally be infinite, we can instead consider the free energy difference between two spacetimes. We consider (2+1)-QFT on an ultrastatic spacetime, so that the free energy is a functional of temperature and the two-space.  
For free scalar and fermion fields in~\cite{FisWalWis18} it was shown that the difference in this free energy for two spaces with the same volume and topology is a physical quantity that is UV finite. Indeed, it may be computed without any regularisation in this free field setting. 
Further, it was shown that the free energy difference for a 2-space that is a perturbation of flat space, relative to a flat space, is both UV and IR finite and quadratic in the amplitude of the metric perturbation, and for any deformation, for all mass and temperature (and scalar curvature coupling for the scalar) the sign of the free energy variation was the same -- flat space is energetically disfavoured. For a general unitary (2+1)-CFT the same energetic revulsion to flat space was shown at zero temperature in \cite{FisWis17}.

In this work we extend the arguments of~\cite{FisWalWis18}. Considering relativistic (2+1)-QFTs on ultrastatic spacetimes we carefully define a free energy difference for arbitrary deformations of flat space relative to flat space itself and argue this is generally UV finite.\footnote{One may do the same for deformations of compact spaces where UV finiteness requires the volume of the deformed and undeformed spaces to be equal.}
For perturbations of flat space, and under reasonable assumptions, it is quadratic in the amplitude of the perturbation and can be computed from the linear response of the one-point function of the stress tensor to the perturbation. As an illustration we use AdS/CFT \cite{Maldacena:1997re,Gubser:1998bc,Witten:1998qj} to compute this for certain strongly coupled theories, holographic CFTs, which have a dual 4-dimensional  gravitational description. AdS/CFT is a powerful tool for study of strongly coupled theories on curved spacetime (see for example the review~\cite{Marolf:2013ioa}) as putting holographic CFTs on a curved space corresponds in the gravity dual to the purely geometric problem of finding Einstein metrics with prescribed conformal boundary. Doing so, and computing the resulting holographic stress tensor one point function following~\cite{Balasubramanian:1999re,Henningson:1998gx,deHaro:2000vlm} we find a similar result to that of the free field theories, namely that the leading variation of free energy decreases for any perturbation and at any temperature. After a suitable normalisation by central charge, there is rather remarkable similarity between the strongly coupled holographic CFT free energy variation and the free fermion CFT (the massless Dirac case computed in~\cite{FisWalWis18}).

In the short wavelength limit (relative to the thermal scale) this perturbative holographic calculation yields the universal zero temperature result for a general CFT in~\cite{FisWis17}. In the opposite long wavelength limit, it can be solved using fluid-gravity methods where the behaviour is governed by hydrodynamics~\cite{Baier:2007ix,Bhattacharyya:2008jc}. More generally we expect at finite temperature in our ultrastatic setting any (2+1)-QFT to have a hydrostatic description. This suggests one may understand the free energy variation as a correction to the ideal fluid stress tensor. We identify the leading correction as a 4 derivative curvature term. In this hydrostatic, or low curvature expansion setting, the free energy difference from flat space may be solved in terms of the integral of the Ricci scalar squared of the deformed space, with a coefficient that in all the theories discussed above has definite sign. This implies that weakly curved two-space is favoured over flat two-space also in the regime where the deformation is not described by a small amplitude metric perturbation. For the free theories discussed in~\cite{FisWalWis18} we explicitly confirm this weak curvature limit which follows simply from the heat kernel expansion~\cite{Vas03} of the determinants that yields the partition function.

The plan for the paper is as follows. In section~\ref{sec:variation} we define the UV finite free energy difference described above for general (2+1)-QFTs. We show that for perturbations of flat space the leading variation in free energy is quadratic in the perturbation amplitude. In section~\ref{sec:review} we briefly review the previous results for this quadratic variation in specific theories.
Then in section~\ref{sec:holoCFT} we compute this quadratic variation for holographic CFTs at  finite temperature which involves computing the boundary stress tensor from linear perturbations of the bulk gravity. Finally in section~\ref{sec:hydro} we derive the fluid-gravity limit for the free energy difference, and then argue that for general (2+1)-QFTs the effect for low curvature deformations of flat space can be understood from hydrostatics, and also confirm these results are true for free fields. 
We conclude with a summary and discussion of the physical interpretation of our free energy difference observable.

\section{Free energy variation}
\label{sec:variation}

We consider a relativistic (2+1)-QFT on a product of time with a static Riemannian 2-space $\Sigma = (\mathcal{M},g_{ij})$, so $d\Sigma^2 = g_{ij}(x) dx^i dx^j$, in the finite temperature thermal vacuum state, with temperature $T$. Moving to Euclidean time, we may regard this quantum thermal system as the QFT on the Riemannian geometry,
\begin{eqnarray}
\label{eq:metric}
ds^2 = \hat{g}_{\mu\nu} dx^\mu dx^\nu = d\tau^2 + g_{ij}(x) dx^i dx^j
\end{eqnarray}
where $\tau \sim \tau + \beta$ with $\beta = 1/T$.\footnote{We are working in $\hbar = c = k_B = 1$ units. }
All (scalar) couplings are constant in spacetime, and  held fixed, and we do not turn on  sources for any non-scalar operators (other than the stress tensor).
The partition function $Z$, which is a functional of $g_{ij}$ and $\beta$,
 defines the thermal vacuum free energy $F$ as,
\begin{eqnarray}
- \beta F = \ln Z[\beta, \Sigma ] \, .
\end{eqnarray}
Let us introduce a UV cut-off $\Lambda$, and then write,
\begin{eqnarray}
Z[\beta, \Sigma] = \int_\Lambda DX e^{- S_E[X; \beta, \Sigma] }
\end{eqnarray}
with $DX$ the integral over fields  (obeying the thermal boundary conditions) and $S_E$ the Euclidean action.
The stress tensor one-point function of this theory in its thermal vacuum, defined as, 
\begin{eqnarray}
\langle T_{\mu\nu} \rangle = - \frac{2}{\sqrt{\hat{g}}} \frac{\delta \ln Z}{\delta \hat{g}^{\mu\nu}}
\end{eqnarray}
is UV divergent without suitable renormalisation.
Since the only inhomogenous deformation of the theory is due to the spatial geometry 
 the divergences
for a diffeomorphism invariant regulator
are given 
by all possible local geometric tensors which are symmetric and conserved and consistent with power counting, which in $(2+1)$-dimensions are a cosmological term and an Einstein tensor term,
\begin{eqnarray}
\langle T_{\mu\nu} \rangle_{\hat{g}} &=& c_1 \Lambda^3 \hat{g}_{\mu\nu} + c_2 \Lambda \hat{G}_{\mu\nu} + O(\Lambda^0)
\end{eqnarray}
with $\hat{G}_{\mu\nu}$ the Einstein tensor, and where $c_{1,2}$
are dimensionless coefficients that depend on the precise theory and its couplings, and nature of the cut-off
and infrared mass scales.
For example, in a theory with a mass $m$, they will be functions going as 
$c_{1,2}(\frac{m}{\Lambda})$
which tend smoothly to a constant as $\Lambda \to \infty$ (keeping temperature and the mass fixed).
The leading divergence going as $\sim \Lambda^3$ leads to the famous `cosmological constant' problem, but there is also the subleading curvature induced divergence going as $\sim \Lambda$   too. 
The stress tensor is renormalised by  suitable local geometric counterterms in the Euclidean action with coefficients that diverge as the cut-off is removed, in this case a cosmological and Einstein-Hilbert term $\int d^3x \sqrt{\hat{g}} \left(  a_1 \Lambda^3 - \frac{1}{2}  a_2   \Lambda \hat{R} \right)$, so that 
$a_{1,2}( \frac{m}{\Lambda})$ are dimensionless functions. In a renormalizable theory, tuning these as $\Lambda \to \infty$ we may then remove the divergences in the stress tensor one-point function (and other correlators) provided the limits, $\lim_{\Lambda \to \infty} \Lambda^3 \left( c_{1} - a_{1} \right)$ and $\lim_{\Lambda \to \infty} \Lambda \left( c_{2} - a_{2} \right)$  exist and are finite. However that leaves a freedom in $a_{1,2}$ corresponding to adding a finite contribution of these counterterms to the action, ie. $a_1 \to a_1 + \alpha_1 \Lambda^{-3}$ and $a_2 \to a_2 + \alpha_2 \Lambda^{-1}$ for any constants $\alpha_{1,2}$. Thus one is left with a finite ambiguity in the stress tensor given by these two local terms.
In curved spacetime QFT, one usually chooses a prescription to ensure that at zero temperature the stress tensor vanishes in flat spacetime, and this 
fixes the finite part associated to the coefficient $a_1$. However, since the Einstein-Hilbert term vanishes in flat space, there is no canonical choice for the finite part of $a_2$. Hence, the renormalised stress tensor, while finite in the $\Lambda \to \infty$ limit, suffers 
ambiguity
parameterised by these pure geometric counterterms in the action.\footnote{
We note that these ambiguities do not arise in the case of a (2+1)-holographic CFT if the theory is only deformed by the metric, as we consider in later section~\ref{sec:holoCFT}. However we emphasize that the discussion above is for a general relativistic (2+1)-QFT.
}
Computing, for example, the energy of a static curved space $\Sigma$, such as a sphere, gives a finite but ambiguous result, which explicitly depends on the nature of the UV physics. 
In such a situation the energy of a given space $\Sigma$ could be arbitrarily negative or positive depending on what finite counterterm contribution the UV theory chooses.
Of course, at finite temperature on a non-compact space, such as the case of deformations of flat space that we are interested in here, the free energy will generally be IR divergent due to the non-zero thermal free energy density being integrated over an infinite volume.

However, as discussed in \cite{FisWalWis18}, we may consider the free energy difference, $\Delta F$, between the theory on the ultrastatic spacetime with compact space $\Sigma$ and a compact reference space $\bar{\Sigma} = (\mathcal{M},\bar{g}_{ij})$ of the same topology and volume, and at the same temperature;
\begin{eqnarray}
- \beta \Delta F[ \beta, \Sigma, \bar{\Sigma} ] = \ln Z[\beta, \Sigma] - \ln Z[\beta, \bar{\Sigma}] 
\end{eqnarray}
As shown in \cite{FisWalWis18} for free scalar and fermion $(2+1)$-dimensional theories this difference  is UV finite, and hence an unambiguous  low energy quantity, independent of details of the UV completion of the theory. Furthermore in the non-compact setting, for  perturbations of flat space, this difference relative to flat space is also IR finite.

We may understand this quantity is UV finite for more general $(2+1)$-dimensional QFTs using the stress tensor divergence structure above.
We begin with the case that our geometries of interest, $\Sigma$ and $\bar{\Sigma}$, are compact with finite volume as this will illustrate the idea. However, we are ultimately interested in the case that $\bar{\Sigma}$ is flat space, and $\Sigma$ is a deformation of it. In this non-compact case there is an added subtlety we shall address after the compact discussion.

Take a smooth one parameter family of (compact) geometries $\Sigma(\epsilon)$ with $\Sigma(0) = \bar{\Sigma}$. In local coordinates the metric on $\Sigma(\epsilon)$ is $g_{ij}(x; \epsilon)$ with $g_{ij}(x; 0) = \bar{g}_{ij}(x)$.
We may define $\Delta F(\epsilon)$ to be the difference of the free energy of $\Sigma(\epsilon)$ to that of $\bar{\Sigma}$.
Then from the definition of the stress tensor, its vev determines the derivative of the partition function and hence the thermal vacuum free energy $F$ as we deform in the parameter $\epsilon$, as
\begin{eqnarray}
\label{eq:varyDF}
\frac{d F}{d\epsilon} =   \frac{1}{2}  \int d^2x \sqrt{{g}} \langle T_{ij} \rangle_{\Sigma(\epsilon)} \frac{d g^{ij}}{d\epsilon} \, .
\end{eqnarray}
The above expression and those that follow are written covariantly in the 2-dimensional geometry unless otherwise explicitly stated.
Here we have assumed that the one-point function of the stress tensor is independent of Euclidean time, as we expect for a good vacuum state, allowing us to  perform the time integral trivially. If this were not the case, the Lorentzian continuation of the vacuum would be time dependent, as could happen for a free tachyonic scalar field.
Substituting our ultrastatic geometry into the general (2+1)-dimensional divergence structure above, we see in our situation of interest the divergence in the spatial components of the stress tensor becomes,
\begin{eqnarray}
\label{eq:divstructure}
\langle T_{ij} \rangle_{\Sigma} &=& c_1 \Lambda^3 g_{ij}  + O(\Lambda^0)
\end{eqnarray}
and there is no contribution from the term involving $c_2$. In the action the corresponding Einstein-Hilbert counterterm for our ultrastatic geometry simply becomes proportional to the Euler characteristic of $\Sigma$, and hence in the variation of $\ln Z$ gives no contribution as the topology of $\Sigma(\epsilon)$ is invariant in $\epsilon$.
Thus, we see,
\begin{eqnarray}
\label{eq:flowdiv}
\frac{d  F}{d\epsilon} = - c_1 \Lambda^3 \frac{d}{d\epsilon} \left(  \int d^2x \sqrt{{g}} \right) + O(\Lambda^0)
\end{eqnarray}
and integrating along the flow,
\begin{eqnarray}
\Delta F = - c_1 \Lambda^3 \left( \mathrm{Vol}(\Sigma) - \mathrm{Vol}(\bar{\Sigma}) \right) + O(\Lambda^0)
\end{eqnarray}
and so provided the volume of the space $\Sigma$, $\mathrm{Vol}(\Sigma)$, and reference geometry, $\mathrm{Vol}(\bar{\Sigma})$, are equal then $\Delta F$ is manifestly finite in the $\Lambda \to \infty$ limit. In computing $\Delta F$ one can use the unrenormalised stress tensor, and clearly the result has no ambiguity
due to the finite part of $a_{1,2}$ cancelling entirely in the difference.

Our focus here will be to consider the reference space $\bar{\Sigma} = (\mathcal{M}, \bar{g})$ to be flat Euclidean 2-space, and choose $\Sigma$ to be a perturbative metric deformation of this. This may be computed as in~\cite{FisWalWis18} by considering a perturbation on a compactified space, such as a torus, and then taking the torus size to infinity keeping the perturbation scale fixed. Alternatively, as we will do later, we may compute directly in the infinite volume setting. Thus, we must now interpret the argument above in this non-compact setting where we must be more careful in handling the infinite volumes of $\Sigma$ and $\bar{\Sigma}$.

We begin in a similar manner, by taking a one parameter family of geometries $\Sigma(\epsilon)$ such that $\Sigma(0) = \bar{\Sigma}$ is flat space. Again, we take local coordinates and write $g_{ij}(x; \epsilon)$ with $g_{ij}(x; 0) = \bar{g}_{ij}(x)$. The subtlety is that the coordinates on the manifold are fixed, and we wish to present both the geometries $\Sigma$ and $\bar{\Sigma}$ in these same coordinates. Thus, we have two metrics and only one coordinate freedom. While the geometry $\bar{\Sigma}$ is fixed as flat space, the explicit metric components may be evolved in the flow parameter $\epsilon$ by a diffeomorphism \emph{relative} to those of $\Sigma$ and this may potentially have physical effect. Thus, we write this metric on $\bar{\Sigma}$ as $\bar{g}_{ij}(x; \epsilon)$ with $\bar{g}_{ij}(x; 0) = \bar{g}_{ij}(x)$. Hence,
\begin{eqnarray}
\frac{d \bar{g}_{ij}(x; \epsilon)}{d\epsilon} = 2 \bar{\nabla}_{(i} v_{j)}
\end{eqnarray}
where $\bar{\nabla}$ is the connection of $\bar{g}_{ij}(\epsilon)$ and $v_i(x,\epsilon)$ is a smooth one parameter family of vector fields that generate the diffeomorphisms on $\bar{g}_{ij}(\epsilon)$ along the flow.

We assume the vacuum on the reference flat space is static so we may use the earlier relation~\eqref{eq:varyDF}. Further we assume the spatial components of the stress tensor one-point function on flat space are simply determined by the homogeneous pressure $p$, so,
\begin{eqnarray}
\langle T_{ij} \rangle_{\bar{\Sigma}} = p \, \bar{g}_{ij}(x;\epsilon)
\end{eqnarray}
at the point $\epsilon$ in the flow. Note if the one-point function is not renormalised, the UV divergence will be the same as in the earlier equation~\eqref{eq:divstructure} so $p = c_1 \Lambda^3 + O(\Lambda^0)$.
Then using equation~\eqref{eq:varyDF} we have,
\begin{eqnarray}
\label{eq:flowbarF}
\frac{d \bar{F}}{d\epsilon} = - p  \frac{d}{d\epsilon} \mathrm{Vol}(\bar{g}, \epsilon) 
\end{eqnarray}
where the volume functional is defined in terms of the metric $\bar{g}_{ij}(x,\epsilon)$ as $\mathrm{Vol}(\bar{g}, \epsilon) =  \int d^2x \sqrt{\bar{g}(\epsilon)}$. Now consider this (divergent) volume functional on the flat space $\bar{\Sigma}$.
In the compact case of course for a given $\bar{\Sigma}$ the volume is fixed and cannot depend on the choice of coordinates. However in this non-compact deformed flat case it may not be fixed if `large diffeomorphisms' are allowed. 
From the definition we see,
\begin{eqnarray}
\frac{d}{d\epsilon} \mathrm{Vol}(\bar{g}, \epsilon) =   \int d^2x \sqrt{\bar{g}(\epsilon)} \bar{\nabla}_i v^i =   \int_{\partial_\infty \mathcal{M}} dS_i v^i 
\end{eqnarray}
with $dS_i$ the outward directed length element of the asymptotic boundary, $\partial_\infty \mathcal{M}$, which is understood as being defined via a suitable limit. If diffeomorphisms are allowed such that this boundary term does not vanish then we see that the variation of $\mathrm{Vol}(\bar{g}, \epsilon)$ in the parameter $\epsilon$ may be finite and non-vanishing. One could disallow such diffeomorphisms, but  this would put an unreasonably strong constraint on the allowed geometries $\Sigma$. 
Hence we learn that the reference flat space  free energy we subtract in the non-compact case may have a coordinate dependence in $\epsilon$, although this is \emph{only} through the variation of volume due to large diffeomorphisms along the flow $\epsilon$.

An explicit example may serve to illustrate this further. Consider the flat reference metric written in polar coordinates, $\bar{g}_{ij}(x) dx^i dx^j= dr^2 + r^2 d\theta^2$. Then we may deform this along the flow by the large diffeomorphism
\begin{eqnarray}
\label{eq:diffeo}
\rho^2 = r^2 \left( \frac{1 + 2 v(\epsilon) r^2 + r^4}{1 + r^4} \right) \; , \quad \bar{g}_{ij}(x,\epsilon) dx^i dx^j= \left( \frac{\partial \rho(r,\epsilon)}{\partial r} \right)^2 dr^2 + \rho(r,\epsilon)^2 d\theta^2 \,,
\end{eqnarray}
provided~$v(\epsilon) > - \frac{4}{3\sqrt{3}}$. This is clearly still flat space, however, if we consider the variation of the volume in $\epsilon$ we find,
\begin{eqnarray}
\frac{d}{d\epsilon} \mathrm{Vol}(\bar{g}, \epsilon) &=&  \frac{d}{d\epsilon} \int_0^\infty dr \int_0^{2\pi} d\theta \rho(r,\epsilon)  \frac{\partial \rho(r,\epsilon)}{\partial r} =   2 \pi \int_0^\infty dr \left( \frac{4 r^3 v'(\epsilon) }{\left( 1 + r^4 \right)^2 } \right) =  2 \pi v'(\epsilon) \, .
\end{eqnarray}
Thus, the coordinate transform $\rho^2 = r^2 + 2 v(\epsilon) + O(\frac{1}{r})$ `stretches' the space $\bar{\Sigma}$ relative to the fixed coordinate chart. Whilst the volume itself is infinite, its variation in $\epsilon$ is finite and non-vanishing.

Now we may proceed as before to consider the UV behaviour, but being careful to note in this non-compact case that the free energy functional evaluated on both~$\Sigma(\epsilon)$ \textit{and} the flat reference space depend on the flow parameter,~$\epsilon$ i.e.  $\beta F(\epsilon) = - \ln Z[ g_{ij}(\epsilon) ]$ and $\beta \bar{F}(\epsilon) = - \ln Z[ \bar{g}_{ij}(\epsilon) ]$ so that $\Delta F(\epsilon) = \Delta F[ \Sigma(\epsilon), \bar{\Sigma} ] = F(\epsilon) - \bar{F}(\epsilon)$.
We have the same expression as previously in equation~\eqref{eq:flowdiv} for the UV divergence of $dF(\epsilon)/d\epsilon$, but now also have a similar expression for $d\bar{F}(\epsilon)/d\epsilon$ leading to,
\begin{eqnarray}
\label{eq:flowdeltaF}
\frac{d \Delta F}{d\epsilon} = - c_1 \Lambda^3 \left( \frac{d}{d\epsilon} \mathrm{Vol}(g, \epsilon) - \frac{d}{d\epsilon} \mathrm{Vol}(\bar{g}, \epsilon) \right) + O(\Lambda^0) \, .
\end{eqnarray}
Whilst the reference geometry is fixed to be flat space we may choose `large diffeomorphisms' to adjust the change in volume $\frac{d}{d\epsilon} \mathrm{Vol}(\bar{g}, \epsilon)$ to equal that of the geometry of interest $\frac{d}{d\epsilon} \mathrm{Vol}(g, \epsilon)$. Doing so then renders $\Delta F(\epsilon)$ to be UV finite. Furthermore since the variation of the reference free energy $\bar{F}(\epsilon)$ only depends on the volume variation, as we saw in equation~\eqref{eq:flowbarF}, this completely fixes the finite part of the reference space subtraction too.
We have seen in the explicit example above that for flat space, by an appropriate choice of the function $v(\epsilon)$ in equation~\eqref{eq:diffeo}, we may always solve this condition (at least in this example, we should be near enough to $\epsilon = 0$ that~$v(\epsilon) > - \frac{4}{3\sqrt{3}}$ remains true, but of course one could use other choices of `large' diffeomorphism). It may be interesting to explore this for other non-compact spaces.

Thus, whilst in the compact case we require the volume of $\Sigma$ and $\bar{\Sigma}$ to be equal to ensure a UV finite free energy difference, the non-compact case is rather different. Due to `large diffeomorphisms'  there is no volume constraint on the geometry $\Sigma$ -- there could not be as the volumes of both $\Sigma$ and $\bar{\Sigma}$ are infinite and not well defined. The key point is that these `large diffeomorphisms' may be used to subtract the `correctly stretched' flat reference geometry, $\bar{\Sigma}$, in order to ensure $\Delta F$ is UV finite. 
Note that had we not stretched the geometry $\bar{\Sigma}$ appropriately, in order to have a UV finite free energy difference we would have to restrict to deformations such that $\frac{d}{d\epsilon} \mathrm{Vol}({g}, \epsilon)$ vanishes, which would be an unreasonably strong restriction on the allowed deformed geometries $g_{ij}(x,\epsilon)$ given a starting flat reference geometry metric $\bar{g}_{ij}(x)$.

Perhaps another more physical way to say this is as follows. Whilst in the compact case one must compare a $\Sigma$ and $\bar{\Sigma}$ with the same volume, one always has the freedom to scale one or other to achieve this volume condition. Such freedom should also be present in the flat non-compact case. Obviously since its volume is infinite, it doesn't make much sense to scale the space, but instead this freedom to `match' the two spaces $\Sigma$ and $\bar{\Sigma}$ appropriately is implemented by these `large diffeomorphisms' or `stretching'. This issue will be further discussed elsewhere~\cite{ongoing}.

In the free field case~\cite{FisWalWis18} one finds that $\Delta F \sim O(\epsilon^2)$ and hence is quadratic in the metric perturbation to flat space, rather than being linear which, naively, one might have expected. We shall now show how to compute $\Delta F$ generally for perturbations of flat space from the variation of the stress tensor, and in particular we shall see why the variation is quadratic.
While we are primarily interested in taking $\bar{\Sigma}$ to be flat space, for the time being we also treat the compact case, which we will see may also lead to the same quadratic behaviour.
Since the spatial geometry is two dimensional, we may choose coordinates so that we  write the deformation of the geometry, $\Sigma(\epsilon)$, as a Weyl deformation of the reference geometry presented as $\bar{g}_{ij}(x)$, so,
\begin{eqnarray}
g_{ij}(x; \epsilon) = e^{2 f(x ; \epsilon)} \bar{g}_{ij}(x)
\end{eqnarray}
where $f$ is a one parameter family of smooth functions on $\bar{\Sigma}$ with $f(x; 0) = 0$ so that $\Sigma(0) = \bar{\Sigma}$. 
We now expand about $\epsilon = 0$ as,
\begin{eqnarray}
f(x) = \epsilon  \, f^{(1)}(x) + \epsilon^2  f^{(2)}(x) + O(\epsilon^3) \, .
\end{eqnarray}
In response to this deformation, we write the perturbation to the vev of the spatial part of the stress tensor due to this metric deformation  as,
\begin{eqnarray}
\langle T_{ij} \rangle_{\Sigma(\epsilon)}  = \bar{\sigma}_{ij}(x) + \epsilon \, \delta \sigma_{ij}(x) + O(\epsilon^2) \, .
\end{eqnarray}
Following the discussion above, one may take this one-point function to be either renormalised or not, as any divergent parts will cancel in the final result.
We view $\delta \sigma_{ij}(x)$ as the linear response of the spatial stress tensor to the metric deformation. Thus, we  think of $\delta \sigma_{ij}$ as a linear functional of $f^{(1)}$, while it is of course independent of the higher orders of the deformation, such as $f^{(2)}$.
For later convenience we denote the energy density $\rho = \langle T_{tt} \rangle_{\Sigma(\epsilon)} = - \langle T_{\tau\tau} \rangle_{\Sigma(\epsilon)}$, and this varies as,
\begin{eqnarray}
\rho  = \bar{\rho} + \epsilon \, \delta \rho(x) + O(\epsilon^2) \, .
\end{eqnarray}

Again we assume the vacuum is static, in the sense that $\langle T_{ij} \rangle_{\Sigma(\epsilon)}$ has only spatial dependence and no dependence on (Euclidean) time. Then we may use the earlier relation~\eqref{eq:varyDF}. We also assume that $\bar{\Sigma}$ has a suitable translation invariance so that $\bar{\sigma}^i_{~i} = \bar{\sigma}_{ij} \bar{g}^{ij} =$constant. 
We would certainly expect this for $\bar{\Sigma}$ being flat spacetime (when $\bar{\sigma}^i_{~i} = 2 p$) but also for other homogeneous (not necessarily isotropic) spaces such as tori and spheres. 
We note that this disallows `striped' phases of vacuum (see for example~\cite{Donos:2011bh}) although in the absence of sources for operators (other than a curved metric) one would not expect the homogeneous vacuum to spontaneously break unless the theory possesses a tachyonic direction which would render it ill-defined on flat spacetime (although perhaps valid on suitably small compact spaces).
Following our assumptions we use equation~\eqref{eq:varyDF} to write,
\begin{eqnarray}
\label{eq:dFdeps}
\frac{d  F}{d \epsilon} & = & - \frac{1}{2} \bar{\sigma}^i_i \left( \frac{d}{d\epsilon} \mathrm{Vol}({g}, \epsilon) \right) + 2 \epsilon \int d^2x \sqrt{\bar{g}}  \left(  \left(  f^{(1)}  \right)^2 \bar{\sigma}^i_{~i}  - \frac{1}{2}  f^{(1)}  \, \delta \sigma^i_{~i}  \right) + O\left( \epsilon^2 \right)
\end{eqnarray}
where indices are raised and lowered using the reference metric $\bar{g}_{ij}(x)$. 
Now in the compact case since we choose the volume of $\Sigma(\epsilon)$ to equal that of $\bar{\Sigma}$ to ensure UV finiteness, so $ \frac{d}{d\epsilon} \mathrm{Vol}({g}, \epsilon) = 0$, then we find,
\begin{eqnarray}
\label{eq:variation}
\Delta F(\epsilon) & = & \epsilon^2 \int d^2x \sqrt{\bar{g}}  \left(  \left(  f^{(1)}  \right)^2 \bar{\sigma}^i_{~i}  - \frac{1}{2}  f^{(1)}  \, \delta \sigma^i_{~i}  \right) + O\left( \epsilon^3 \right) \, .
\end{eqnarray}
However, following our discussion above we obtain precisely the same expression in the non-compact case for deformations of flat space, since for UV finiteness we choose appropriate `large diffeomorphisms' for $\bar{g}_{ij}(x,\epsilon)$ so that $\frac{d}{d\epsilon} \mathrm{Vol}(\bar{g}, \epsilon)$ equals $\frac{d}{d\epsilon} \mathrm{Vol}(g, \epsilon)$, and $\bar{F}(\epsilon)$ evolves as in equation~\eqref{eq:flowbarF}.
Hence, for deformations of both homogeneous compact spaces and flat space, where the vacuum of the undeformed space is static and has constant $\bar{\sigma}^i_i$,
we arrive at a quadratic variation of our free energy difference. This is determined in terms of the linear deformation of the metric, $f^{(1)}$, both explicitly and implicitly through the response of the spatial stress components $\delta \sigma_{ij}$. If we chose to use the unrenormalised stress tensor one-point function, from earlier equation~\eqref{eq:divstructure} we would find that $\bar{\sigma}^i_{~i} = 2 c_1 \Lambda^3 + O(\Lambda^0)$ and $\delta {\sigma}^i_{~i} = 4 c_1 \Lambda^3 f^{(1)} + O(\Lambda^0)$, and hence the UV divergences will cancel between the two terms above, leaving only a UV finite result as expected.
Note that if one chose a reference geometry which was not homogeneous, and hence presumably $\bar{\sigma}^i_i$ would not be constant, then one would expect a linear variation in $\epsilon$ instead. It is the quadratic nature of the variation for homogeneous  spaces that potentially enables $\Delta F$ to have a definite sign.

We now specialise to the case of the reference space $\bar{\Sigma}$ being flat space, and choose natural coordinates so that $\bar{g}_{ij} = \delta_{ij}$.
Then we may decompose the leading metric perturbation as a Fourier transform,
\begin{eqnarray}
\label{eq:Fourierf}
f^{(1)}(x)  & = & \int d^2k \, e^{ i k_i x^i} \tilde{f}(k_i) 
\end{eqnarray}
where reality imposes $\tilde{f}(- k_i) =   \tilde{f}(k_i)^\star$. On flat space the linear response of the trace of the spatial stress tensor, $\delta \sigma^i_i$, to the metric deformation is constrained by the rotational and translation invariance. For a deformation by a single Fourier mode the response will be proportional to that mode, with a coefficient depending on the wavevector $k_i$ only through its  magnitude, $k = \sqrt{k_i k_i}$. Hence, for a general perturbation the response will be
\begin{eqnarray}
\label{eq:Fouriersigma}
\delta \sigma^i_i(x) & = & \int d^2k \, e^{ i k_i x^i} s(k) \tilde{f}(k_i)
\end{eqnarray}
and is characterised by the  function $s(k)$. Then 
 we may write the quadratic variation of $\Delta F$ as,
\begin{eqnarray}
\label{eq:quadvary}
\Delta F(\epsilon) & = & - \epsilon^2 \int d^2k \, a( k ) \left| \tilde{f}(k_i) \right|^2 + O\left( \epsilon^3 \right)
\end{eqnarray}
where the function $a(k) = (2\pi)^2 \left(\frac{1}{2} s(k) - \bar{\sigma}^i_i \right)$ characterises the variation, and again only depends on the wavevector through its magnitude $k$. Note that as we have defined signs, modes which have positive $a(k)$ give rise to a quadratic \emph{decrease} in free energy relative to flat space.

\section{Review: zero temperature CFT and finite temperature free fields}
\label{sec:review}

We now review the results of this free energy variation given in previous computations.
Firstly let us consider a general CFT at zero temperature, so we have an energy variation rather than free energy variation with metric deformation which we may interpret as a vacuum Casimir energy. We define the CFT `central charge' $c_T$ as the coefficient entering the two point function of the stress tensor in vacuum on flat space as,
\begin{eqnarray}
\langle T_{\mu\nu}(x) T_{\alpha\beta}(0) \rangle  & = & \frac{c_T}{|x|^6} \left( I_{\mu(\alpha} I_{\beta)\nu} -\frac{1}{3} \delta_{\mu\nu} \delta_{\alpha\beta} \right)
\end{eqnarray}
with $I^{\mu\nu} = \delta^{\mu\nu} - 2 x^\mu x^\nu / |x|^2$,
and for a unitary theory $c_T > 0$ \cite{OsbornPetkou}. Then in~\cite{FisWis17} it was shown that the energy variation is as above with a positive function $a(k)$ given as,
\begin{eqnarray}
 a_{\mathrm{CFT}}( k ) & = & c_T \frac{\pi^4}{24} k^3 > 0
\end{eqnarray}
leading to flat space being disfavoured over any metric deformation of it at zero temperature. In fact for a holographic CFT at zero temperature it was shown in~\cite{FisWis17} using the dual gravitational methods of~\cite{HicWis15} that for non-perturbative deformations of flat space the energy is negative.

Secondly, consider either a free scalar $\phi$, or a Dirac fermion $\psi$, and take the field to have mass $M$ (which may be zero). In the scalar case we include a scalar curvature coupling $\xi$. Thus, the fields obey the (2+1)-dimensional equations of motion,
\begin{eqnarray}
\left(-\hat{\nabla}^2 + \xi \hat{R} + M^2 \right) \phi = 0 \, , \quad \left(\hat{\slashed{D}} + M \right) \psi = 0 
\end{eqnarray}
where~$\hat{\slashed{D}}$ is understood as being defined by the spacetime spin connection. 
Then the variation of free energy at fixed finite temperature $T$ for a perturbation to flat space was determined in~\cite{FisWalWis18} using heat kernel methods to be,
\begin{eqnarray}
\frac{d  F}{d \epsilon} & = & - 2 \epsilon \int d^2k \, b_{\mathrm{s},\mathrm{f}}( k ) \left| \tilde{f}(k_i) \right|^2 + O\left( \epsilon^3 \right)
\end{eqnarray}
where the perturbation function $f(x)$ was constrained to leave the volume $\mathrm{Vol}(\Sigma)$ invariant, so $\int d^2x f^{(1)}(x) = \int d^2x \left( f^{(2)}(x) + \frac{1}{2} f^{(1)}(x)^2 \right) = 0$, but was otherwise arbitrary. In particular this implies $\tilde{f}( 0_i ) = 0$.
The functions $b_{\mathrm{s},\mathrm{f}}( k )$ were found to be,
\begin{eqnarray}
\label{eq:adeffree}
b_{\mathrm{s},\mathrm{f}}(k) = -q T k^4 \int_0^\infty dt \, e^{-M^2 t} \Theta_q(T^2 t) I_{\mathrm{s},\mathrm{f}}(k^2 t).
\end{eqnarray}
where the thermal factor is given as,
\begin{eqnarray}
\label{eq:Thetadef}
\Theta_q(\zeta) = \sum_{n=-\infty}^{\infty} e^{- (2 \pi)^2 (n-q+1/2)^2 \zeta}
\end{eqnarray}
and for a scalar we take $q = - \frac{1}{2}$ and
\begin{eqnarray}
\label{eq:scalar}
I_\mathrm{s}(\zeta) &=& -\frac{\pi}{4 \zeta^2} \left[6 +  \zeta ( 1 - 8 \xi)  - \left( 6 + 2 \zeta (1 - 4 \xi) + \frac{\zeta^2}{2}  (1 - 4 \xi)^2 \right) \mathcal{F}\left( \frac{\sqrt{\zeta}}{ 2} \right) \right]
\end{eqnarray}
whereas for the fermion we take $q = +1$ and 
\begin{eqnarray}
\label{eq:dirac}
I_\mathrm{f}(\zeta) = \frac{\pi}{4 \zeta^2} \left[\left(6 + \zeta\right)\mathcal{F}\left(\frac{\sqrt{\zeta}}{ 2} \right) - 6 \right] \, .
\end{eqnarray}
Here we have defined~$\mathcal{F}(\zeta) = \zeta^{-1} e^{-\zeta^2} \int_0^{\zeta} d\zeta' \, e^{(\zeta')^2}$.

While $f(x)$ here was constrained to leave volume invariant, but otherwise arbitrary, by comparison with equation~\eqref{eq:dFdeps} for a general perturbation which may vary volume we see that $a(k) = b_{\mathrm{s},\mathrm{f}}( k )$. This is true for all non-zero $k$ by direct comparison, and since $a(k)$ should be smooth in the limit $k \to 0$ (since the theory is local, and this is the long wavelength limit) it must also hold at $k=0$.
As discussed in~\cite{FisWalWis18} in both scalar and fermion cases then the function $a(k) > 0$, and hence perturbed flat space is energetically preferred  relative to flat space itself, for any mass $M$ (including zero) and for any temperature $T$ (including zero), and in the case of the scalar, for any curvature coupling $\xi$.

\section{Finite temperature holographic CFT}
\label{sec:holoCFT}

We now consider computing the leading free energy difference at finite temperature for a holographic CFT deforming away from a flat reference space. 
From the above discussion we may compute this  using equation~\eqref{eq:quadvary} by considering the linear response of the spatial components of the stress tensor, $\delta \sigma_{ij}$, to the linear perturbation $f^{(1)}$ to the flat space the theory is defined on. 

We make the assumption that the behaviour of the holographic CFT is governed by only the `universal sector', and hence is computed from a dual 4-dimensional gravity with a negative cosmological constant. Since we are only deforming by turning on temperature and changing the spatial geometry perturbatively from flat, with no other sources, we expect that even at finite temperature the thermal vacuum is described by this universal sector. This would of course be different if other sources, such as chemical potentials were turned on, which could induce finite temperature phase transitions involving condensing fields outside the universal sector (as for example famously for holographic superconductors~\cite{Gubser:2008px,Hartnoll:2008vx,Hartnoll:2008kx}). This could also differ if one perturbed a different homogeneous boundary space, such as a round sphere where the dual global AdS Schwarzschild would be unstable at low temperature to localisation on an internal space~\cite{Gregory:1993vy,Hubeny:2002xn,Dias:2015pda,Dias:2016eto}.

The thermal vacuum solution dual to the holographic CFT on the reference flat space boundary is then planar AdS-Schwarzschild, which we write in Euclidean time as,
\begin{eqnarray}
ds^2_{(\mathrm{bulk})} &=& g^{(\mathrm{Sch})}_{AB} dx^A dx^B \nonumber \\ 
& = & \frac{\ell^2}{z^2} \left( \left( 1 - \left( \frac{z}{z_0} \right)^3 \right) d\tau^2 + dx_i^2 + \frac{dz^2}{ 1 - \left( \frac{z}{z_0} \right)^3} \right)
\end{eqnarray}
where temperature $T = \frac{3}{4 \pi  z_0}$, and this solves the bulk Einstein equation $R_{AB} = - \frac{3}{\ell^2} g_{AB}$, where the AdS-scale $\ell$ and bulk Newton constant $G_N$ are related to the dual CFT `central charge', $c_T^{(\mathrm{holo})}$, as $\frac{\pi^2}{48} c_T^{(\mathrm{holo})} = \frac{\ell^2}{16\pi G_N}$ (see for example \cite{Liu:1998bu,Buchel:2009sk}). The spatial part of the boundary metric is the flat reference metric, $\bar{g}_{ij} = \delta_{ij}$, in the coordinates $x^i$. The (traceless) stress tensor vev for this reference geometry is constant with,
\begin{eqnarray}
  \bar{\sigma}_{ij} =  \frac{4 \pi^5}{81} T^3 c_T^{(\mathrm{holo})}  \delta_{ij} 
\end{eqnarray}
so that ${\rho} =  \bar{\sigma}^i_i = \frac{8 \pi^5}{81} T^3 c_T^{(\mathrm{holo})}$.
Following the discussion above we perturb the boundary metric to $g_{ij}(x ; \epsilon) = \left( 1 + 2 \epsilon\, f^{(1)}(x) + O(\epsilon^2) \right) \delta_{ij}$ and expect a quadratic free energy response as in equation~\eqref{eq:quadvary}.
 To leading order in $\epsilon$ this results in a bulk solution which is a linear deformation to the homogeneous black hole,
\begin{eqnarray}
ds^2_{(\mathrm{bulk})} = \left( g^{(\mathrm{Sch})}_{AB} + \epsilon \, h_{AB} + O(\epsilon^2) \right) dx^A dx^B
\end{eqnarray}
with the prescribed spatial boundary geometry $\Sigma(\epsilon)$ with metric $g_{ij}(x ; \epsilon)$, together with regularity at the bulk thermal horizon giving boundary conditions for the perturbation $h_{AB}$. This linear bulk perturbation  induces a linear variation $\delta \sigma_{ij}$ in the boundary stress tensor. Thus, our task is simply to find the appropriate linear bulk perturbation.

Before we proceed a comment is in order. Since the CFT partition function can be computed from the renormalised on-shell action when the theory is described by a gravity dual, one might think that one should  directly compute this on-shell action to yield $\Delta F$, rather than going via computation of the stress tensor. While one could do this, we know that since the variation is quadratic in the perturbation, then one would naively have to work to second order in $\epsilon$ in the bulk which is a considerably more complicated task. In fact we have done the calculation this way too in the same fashion as for the perturbative construction of nonuniform black strings \cite{Gubser:2001ac,Horowitz:2011cq}, and of course it agrees as it must, so we will not detail it here. The important point we emphasise is that the first law like relation in equation~\eqref{eq:variation} constrains the variation $\Delta F$ and shows us that really the result only depends on the perturbation and its response at linear order, even though the resulting $\Delta F$ is quadratic.

Consider for now the bulk response to a single mode of the boundary deformation,
\begin{eqnarray}
\label{eq:singlemode}
 f^{(1)}(x^i) & = & e^{i k x}
\end{eqnarray}
in the $x$ direction. Since the perturbation is trivial if $f^{(1)}$ is constant, we shall assume that $k \ne 0$.
The symmetry in the $\tau$ and $y$ directions implies the bulk perturbation takes the general form,
\begin{eqnarray}
 h_{AB} & = & \frac{\ell^2 e^{i k x}}{z^2} \left( 
\begin{array}{cccc}
 h_{\tau\tau}(z) & 0 & 0  & 0 \\
&  h_{zz}(z) & h_{xz}(z) & 0 \\
& h_{xz}(z) & h_{xx}(z) & 0 \\
& & &  h_{yy}(z)
\end{array}
\right) 
\end{eqnarray}
writing $x^A = ( \tau, z, x, y )$. Considering a general diffeomorphism, $v^A = ( 0, a(z) e^{i k x},  i k b(z) e^{i k x}, 0)$, of the Schwarzschild background, then
\begin{eqnarray}
 \nabla^{(\mathrm{Sch})}_{~~~(\tau} v^{}_{\tau)}  =   - \left( 1 + \frac{1}{2} \left( \frac{z}{z_0} \right)^3 \right) a(z) \frac{\ell^2 e^{i k x}}{z^3} \; , \quad
  \nabla^{(\mathrm{Sch})}_{~~~(x} v^{}_{x)} -   \nabla^{(\mathrm{Sch})}_{~~~(y} v^{}_{y)}  =  - k^2 b(z) \frac{\ell^2 e^{i k x}}{z^2}
 \end{eqnarray}
and hence (noting that $k \ne 0$) we may completely fix the gauge for the perturbation with the choice $h_{\tau\tau}(z) = 0$ and $h_{xx}(z) = h_{yy}(z)$. 
Defining the function $2 u(z) = h_{xx}(z) = h_{yy}(z)$ one finds that Einstein's equation reduces to the single second order o.d.e., most conveniently written defining the variable $\chi = z/z_0$ and dimensionless wavevector $\tilde{k} = z_0 k$,
\begin{eqnarray}
\label{eq:ueqn}
u''(\chi) -  \left( \frac{16 - 24 \chi^3 + 36 \chi^6 - \chi^9}{8 - 6 \chi^3 - 3 \chi^6 + \chi^9} \right) \frac{ u'(\chi) }{\chi}  - \frac{\tilde{k}^2}{1 - \chi^3} u(\chi) & = & 0 
\end{eqnarray}
with the remaining non-zero metric components being given in terms of the solution as,
\begin{eqnarray}
h_{zz}(z) = -   \frac{z}{ 1 - \frac{ \chi^3}{4}}  u'(z) \; , \quad
h_{xz}(z) = - \frac{2 i}{k} \left( \frac{2 +  \chi^3}{ 4 -  \chi^3 } \right)^2 u'(z) -  \frac{2 i k z }{ 4 -  \chi^3 } u(z) \, .
\end{eqnarray}
We now consider the two boundary conditions on $u(\chi)$ required by this second order o.d.e.

Performing an asymptotic expansion about $z = 0$ we find the general solution for $u$ behaves as a power series,
\begin{eqnarray}
u = u_0 - \ \tilde{k}^2 u_0 \frac{\chi^2}{2} +  u_3 \frac{\chi^3}{6} + O(\chi^4)
\end{eqnarray}
where higher powers of $\chi$ have their coefficients determined in terms of the two constants of integration for the o.d.e. $u_0$ and $u_3$.
In order that the boundary metric takes the required form corresponding to the single mode perturbation in equation~\eqref{eq:singlemode} we must impose the condition $u_0 = 1$. 
The coefficient $u_3$ is determined by the requirement that the horizon remains regular. In our gauge choice the horizon remains at $z = z_0$ (or $\chi = 1$), where the two behaviours are the smooth one we require which is a Taylor series in $(1-\chi)$,
\begin{eqnarray}
u = u_h +  O\left( 1- \chi \right)
\end{eqnarray}
with the coefficient of higher powers of $(1-\chi)$  given in terms of the value of $u$ at the horizon, $u_h$, and also an unwanted singular behaviour $u(z) \sim u_l \log\left( z_0 - z \right)$. The general solution is the sum of these, and our boundary condition is to ensure $u_l = 0$, which in turn will determine the asymptotic data $u_3$. An important point is that the regular behaviour does not perturb the surface gravity, and hence the temperature, at linear order in $\epsilon$ -- this can be understood due to this mode having non-trivial spatial variation in the horizon directions, whereas surface gravity must be constant over the horizon.

Thus, imposing the boundary conditions determines $u_3$ as a function of $k$ and $z_0$, which due to scale symmetry is only a function of the dimensionless $\tilde{k}$. Let us denote this solution as $u_3 = u_3(\tilde{k})$.
We are unable to determine this analytically, but it is a straightforward numerical exercise to compute it using standard `shooting' methods and we will discuss the full solution shortly.
The low temperature limit, $\tilde{k} \to \infty$, is simple to obtain. Then the linear equation~\eqref{eq:ueqn} for $u$ reduces to,
\begin{eqnarray}
\label{eq:ueqnlowT}
u''(z) -  \frac{2}{z}  u'(z)   - k^2 u(z) & = & 0 
\end{eqnarray}
with solution, $u(z) = e^{- k z} (1 + k z)$ obeying the boundary conditions and implying $u_3(\tilde{k}) = 2 \tilde{k}^3$ for large $\tilde{k}$. 

Having determined $u_3(\tilde{k})$ from the o.d.e.~\eqref{eq:ueqn} and its boundary conditions, we may  deduce the linear variation of the holographic CFT stress tensor.
We move to a Feffermann-Graham gauge transforming from $z, x$ to new coordinates $z' = z_0\chi'$, $x'$ so that,
\begin{eqnarray}
ds^2_{(\mathrm{bulk})} & = & \frac{\ell^2}{z'^2} \left( dz'^2 + g'_{\tau\tau} d\tau^2 + g'_{ij} dx'^i dx'^j \right) 
\end{eqnarray}
which is achieved by taking,
\begin{eqnarray}
\chi &=& \chi' \left( 1 - \frac{1}{6} \chi'^3 + O(\chi'^5) \right) + \epsilon  e^{i k x'} \left( - \frac{\tilde{k}^2}{4} \chi'^3 + \frac{u_3}{12} \chi'^4 + O(\chi'^5) \right) \nonumber \\
x & = & x' + \frac{\epsilon}{i k} e^{i k x'}\left( - \frac{u_3}{12} \chi'^3 + \frac{\tilde{k}^4}{16} \chi'^4 + O(\chi'^5) \right)
\end{eqnarray}
so that near the conformal boundary the metric is given in terms of the CFT stress tensor as \cite{Balasubramanian:1999re,Henningson:1998gx,deHaro:2000vlm},
\begin{eqnarray}
g'_{\tau\tau} & = & 1 + \epsilon \frac{{k}^2}{2}  e^{i k x} z'^2 -  \frac{1}{3 c_T^{(\mathrm{holo})}}  \left( \bar{\rho} + \epsilon \delta \rho(x') \right) z'^3 + O(z'^4) \nonumber \\
g'_{ij} & = &  \left( 1 + 2 \epsilon e^{i k x} \right) \delta_{ij} - \epsilon \frac{{k}^2}{2}  e^{i k x}  z'^2 + \frac{1}{3 c_T^{(\mathrm{holo})}} \left( \bar{\sigma}_{ij} + \epsilon   \delta \sigma_{ij}(x') \right) z'^3 + O(z'^4) 
\end{eqnarray}
where the linear response, $\delta{\sigma}_{ij}$, of the spatial stress tensor is given as,
\begin{eqnarray}
\label{eq:sigmaexpr}
 \delta \sigma_{xx}(x)  & = & \frac{\pi^2 }{24 z_0^3} e^{i k x} c^{(\mathrm{holo})}_T
 \nonumber \\
\delta \sigma_{yy}(x)  & = &  \frac{\pi^2 }{24 z_0^3}  \left(1 + \frac{1}{4} u_3(\tilde{k}) \right)  e^{i k x} c^{(\mathrm{holo})}_T
\end{eqnarray}
with $\delta \sigma_{xy} = 0$ and $\delta \rho = \delta \sigma_{ii}$. 
Using linearity and rotational invariance then for a general Fourier decomposed perturbation of the boundary as in~\eqref{eq:Fourierf}  we will find the trace,
\begin{eqnarray}
\label{eq:sigmatr}
 \delta \sigma^i_i(x)  & = &  \frac{\pi^2 }{24 z_0^3}  c^{(\mathrm{holo})}_T \int d^2k \, \tilde{f}(k_i)  \left( 2 +  \frac{1}{4} u_3(z_0 k) \right) e^{i k_i x^i}  
\end{eqnarray}
and using equation~\eqref{eq:quadvary} this determines the quadratic variation of the free energy, $\Delta F$, in terms of the function $a(k)$ characterising it as,
\begin{eqnarray}
a(k , T) & = &  \frac{ \pi^4  k^3}{48} c^{(\mathrm{holo})}_T    \left( \frac{3 k}{4 \pi T} \right)^{-3} u_3\left( \frac{3 k}{4 \pi T} \right)
\end{eqnarray}
where we have used the relation between $T$ and $z_0$. 
Recall in the low temperature limit, $u_3 \simeq 2 \tilde{k}^3$, which implies $a(k)  = \frac{c^{(\mathrm{holo})}_T \pi^4}{24} k^3$ at low temperature which reproduces the zero temperature general CFT result of \cite{FisWis17} as it should.

\begin{figure}
\centering
\includegraphics[width=0.8\textwidth,page=1]{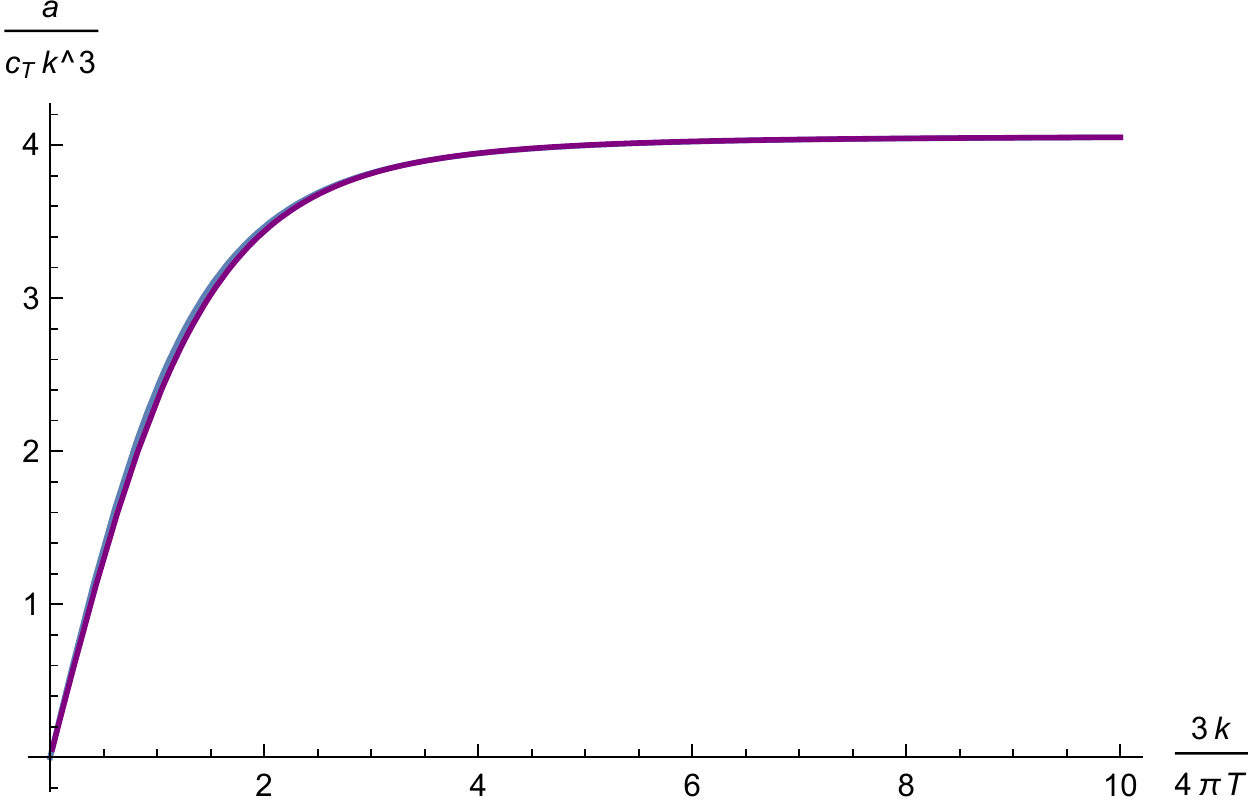}
\caption{Graph of the function $a(k, T)$ which specifies the quadratic variation of the free energy difference from flat space, $\Delta F$, normalised by the CFT central charge, $c_T$, and by $k^3$, against $3 k / (4 \pi T)$ for both the holographic CFT (red curve) and free fermion CFT (blue curve).
For any CFT this asymptotes to $\pi^4/24$ for large $k/T$. However, remarkably both these CFTs have strikingly similar curves over the range of $k/T$ even though they are very different theories.
Since we see $a(k, T)$ is positive for a holographic CFT, this implies that flat space is energetically disfavoured over any perturbation of it, and for any temperature, for such theories.
}
\label{fig:acurves}
\end{figure}

We now turn to the solution, $u_3$, of the numerical shooting problem which directly determines $\Delta F$. In figure \ref{fig:acurves} we plot $\frac{a(k,T)}{c_T k^3}$ for the holographic CFT given by this numerical solution  as a function of the dimensionless variable $\frac{3}{4 \pi} \frac{k}{T}$. We see this tends to $\frac{\pi^4}{24}$ as expected at low temperature. A key observation is that $u_3$, and hence the function $a$, is positive for all $k/T$. This results in the quadratic variation of $\Delta F$ for this strongly coupled holographic CFT being negative for any perturbation of flat space, at any temperature, giving the same qualitative behaviour as for the free fields.
It is very interesting to compare the free energy variation for the holographic CFT to the free fermion CFT, given by the massless Dirac case above. Using the value for the central charge of the massless Dirac fermion $c^{(\mathrm{Dirac})}_T = 3/(4\pi)^2$, we also plot in the same figure $\frac{a(k,T)}{c_T k^3}$ computed by numerically evaluating the integral in equation~\eqref{eq:adeffree} with integrand determined from equation~\eqref{eq:dirac}.
The holographic CFT and free fermion give such similar behaviour of $\Delta F$ when normalised by their central charges that it is hard  to see by eye the two separate curves.\footnote{
We do not compare to the massless conformally coupled scalar, since whilst this is formally a CFT, as mentioned in \cite{FisWalWis18}, the flat space zero mode of the massless scalar with any non-zero curvature coupling   becomes a tachyonic mode when any curvature is present and hence changes the theory to be unstable in character.
}
One might be suspicious that since both curves are determined numerically perhaps they are actually the same. This is not the case, and we shall see shortly that the low wavenumber behaviour confirms the curves cannot be identical.
 That said, we have no idea why these curves are so similar -- presumably it is a coincidence as the theories are quite different, for example one being a free theory with one field and the other strongly coupled with a large number of degrees of freedom.

\section{Hydrostatics and the long wavelength limit}
\label{sec:hydro}

We now investigate the hydrostatic long wavelength limit, $k \to 0$ at fixed $T$ for the holographic theory. As we shall see, this may be viewed as the fluid-gravity limit~\cite{Baier:2007ix,Bhattacharyya:2008jc} which may be applied in the case of weakly deformed boundary metrics~\cite{Bhattacharyya:2008mz,Bhattacharyya:2008mz}. Following the usual procedure we consider the solution in the small $k$ limit relative to the temperature, or equivalently $z_0$, and hence we are working in the limit $\tilde{k} \to 0$. Then we expand the solution as,
\begin{eqnarray}
u(z, k) = u^{(0)}(\chi) + \tilde{k}^2 u^{(2)}(\chi) + \tilde{k}^4 u^{(4)}(\chi) + O(\tilde{k}^6) 
\end{eqnarray}
and solving the linear o.d.e. in equation~\eqref{eq:ueqn} order by order in $\tilde{k}$, and imposing the boundary conditions, one finds,
\begin{eqnarray}
u^{(0)}(\chi) & = & 1 \nonumber \\
u^{(2)}(\chi) & = & - \frac{\chi^2}{2 + \chi^3} \nonumber \\
u^{(4)}(\chi) & = & \frac{1}{108} \left( -2  \sqrt{3} \pi + \frac{36 ( \chi + \chi^3)}{2 + \chi^3} - 9 \log\left( 1 + \chi + \chi^2 \right) +   \,6 \sqrt{3} \arctan\left( \frac{\sqrt{3}}{1+2 \chi} \right) \right) \, .
\end{eqnarray}
From this solution we deduce the data $u_3(\tilde{k}) = \frac{4}{3} \tilde{k}^4$ and so,
\begin{eqnarray}
\label{eq:longwaveholo}
a(k, T) & = &    c^{(\mathrm{holo})}_T  \frac{\pi^3 k^4}{48  T} + O(k^6)
\end{eqnarray}
which accounts for the linear behaviour in the figure \ref{fig:acurves} as $k/T \to 0$. Using equations~\eqref{eq:sigmaexpr} we may read off the CFT stress tensor, raising the question what corrections to the leading perfect conformal fluid stress tensor are responsible for this behaviour. Since the stress tensor perturbation goes as $\sim k^4$ we see that this must involve 4 derivative corrections, and in particular these must be curvature corrections since the effect results from spatial deformations.

More generally in any (2+1)-QFT one can ask whether the leading quadratic variation of the free energy we have considered here can be understood in terms of hydrostatic stress tensor corrections at finite temperature in the long wavelength limit (relative to the thermal scale). Consider our free scalar and fermion theories.
From equation~\eqref{eq:adeffree} we see at long wavelengths compared to $T^{-1}$ and $M^{-1}$,
\begin{eqnarray}
a_{\mathrm{s},\mathrm{f}}(k,T) = - q  I_{\mathrm{s},\mathrm{f}}(0) T k^4 \int_0^\infty dt \, e^{-M^2 t} \Theta_q(T^2 t) + O(k^6)
\end{eqnarray}
which we may evaluate in both cases, giving for the scalar,
\begin{eqnarray}
\label{eq:longwavescalar}
a_{\mathrm{s}}(k,T) = \frac{\pi}{60  M} \left( 1 - 10 \xi + 30 \xi^2 \right)  \coth\left( \frac{M}{2 T} \right) k^4 + O(k^6)
\end{eqnarray}
and for the fermion,
\begin{eqnarray}
\label{eq:longwavefermion}
a_{\mathrm{f}}(k,T) =   \frac{ \pi}{120   M}  \tanh\left( \frac{M}{2 T} \right) k^4 + O(k^6)\, .
\end{eqnarray}
Thus, again we  see the same behaviour $a(k, T) \sim k^4$ as $k \to 0$ at fixed temperature for the fermion, and for the massive scalar, suggesting  the free energy difference may have a long wavelength description in terms of curvature corrections to a fluid stress tensor.

We pause to  emphasise that this relatively fast decay $a(k, T) \sim k^4$ at long wavelength allows very generous asymptotic behaviour of perturbations of flat space that still yield a finite free energy difference. One might have thought  that the perturbation $f$ would be required to have some power law fall-off  to ensure an IR finite free energy difference. In fact we see from equation~\eqref{eq:quadvary} that this is not the case, and a suitably smooth bounded $f$ will give an IR finite result.
We note that for the massless scalar the behaviour is different due to the Euclidean zero mode that theory possesses on flat space which alters the long wavelength behaviour so that $a \sim k^2$ for zero curvature coupling, $\xi$, and as noted earlier, the theory is pathological for non-zero $\xi$.
We also note that for the free fermion CFT, the massless free fermion case, we obtain,
\begin{eqnarray}
\label{eq:longwavediracCFT}
a_{\mathrm{f,CFT}}(k,T) & = & c^{(\mathrm{Dirac})}_T  \frac{\pi^3}{45 T} k^4 + O(k^6)
\end{eqnarray}
when writing it in terms of the free fermion central charge. Thus we see from equations~\eqref{eq:longwaveholo} and~\eqref{eq:longwavediracCFT} that the quantity $a/c_T$ is very similar for the holographic CFT and the free fermion, but is not identical, having the ratio $45/48 \simeq 0.94$ at long wavelengths. This confirms our earlier assertion that the two curves in figure \ref{fig:acurves}, whilst remarkably similar, are not identical. It is also peculiar that the transcendental structure of this limit of the function $a(k,T)$ is the same for the holographic and free fermion CFT.

We now identify the 4 derivative correction to the perfect fluid stress tensor responsible for this leading long wavelength free energy variation for a general (2+1)-dimensional QFT. We assume the theory admits a hydrostatic description at finite temperature for weak deformations of the spatial geometry from flat space. From gauge-gravity duality we know this to be true for the holographic CFT, which admits a hydrodynamic description for weak curvatures \cite{Bhattacharyya:2008ji,Bhattacharyya:2008mz}. From the discussion above it appears to hold  for free scalars and fermions (except in the massless scalar case) as suggested by the behaviour $a(k,T) \sim k^4$, and we will shortly confirm this explicitly. Note that normally one would expect some interactions to be necessary in order to discuss hydrodynamics. While this is true in a dynamical setting, in the canonical equilibrium setting of hydrostatics where the temperature is maintained by an external bath, one can perfectly well consider the fluid of thermal free particles. The dynamical zero mode of the massless scalar theory appears not to have such a local particle interpretation, and modifies this expectation. 

To proceed we consider the most general relativistic (2+1)-dimensional fluid stress-energy tensor $T_{\mu\nu}$ on our ultrastatic curved spacetime up to 4 derivatives in the absence of any other sources or currents. We assume the theory is such that on flat space the thermal vacuum  stress-energy tensor is static and homogeneous. 
Let us for a moment switch to Lorentzian signature.  A dynamical fluid is described by a local temperature $T(x)$ and velocity $u^\mu(x)$ from which the stress tensor is composed as a derivative expansion, the equations of motion of the fluid following from its conservation (see for example~\cite{Rangamani:2009xk} for a nice review of hydrodynamics). 
Then there are two key simplifications in our hydrostatic setting. Firstly since the fluid is static after the spatial deformation, the fluid vector is simply $u = \partial/\partial t$, and in our geometry ${\nabla}_\mu u_\nu = 0$ and hence no terms involving derivatives of the velocity can contribute to the stress tensor. For example, the leading corrections to the perfect fluid stress-energy tensor due to viscosity vanish for our static situation. Likewise since we are considering the thermal vacuum at constant temperature the local temperature function is constant, and no temperature derivative terms will contribute.
Since we have no other sources, and velocity and temperature derivatives vanish, the only relevant derivative corrections are those arising from spacetime curvatures as we should expect since our effect is driven by spatial curvature.  Then the terms that can contribute to $\langle T_{\mu\nu} \rangle$ for our deformation of the flat  geometry, and are both compatible with symmetry and conservation, are then,
\begin{eqnarray}
\langle T_{\mu\nu} \rangle_\Sigma &=&  p \, \hat{g}_{\mu\nu}  + a \left( \hat{R}_{\mu\nu} - \frac{1}{2} \hat{R} g_{\mu\nu} \right) + b B_{\mu\nu} + c C_{\mu\nu}  \nonumber \\
&& + u_\mu u_\nu \left( ( \rho + p )  + d \hat{R} + e \hat{\nabla}^2 \hat{R}  + f \hat{R}^2 + g ( \hat{R}_{\alpha\beta} )^2 \right) \nonumber \\
&& + \ldots
\end{eqnarray}
where this expression is written covariantly in the full (2+1)-spacetime, and $B_{\mu\nu}$ and $C_{\mu\nu}$ are the two conserved symmetric 4-derivative tensors,
\begin{eqnarray}
B_{\mu\nu} & = & \hat{\nabla}_\mu \partial_\nu \hat{R} - \hat{\nabla}^2 \hat{R} \hat{g}_{\mu\nu} - \hat{R} \hat{R}_{\mu\nu} + \frac{1}{4} \hat{R}^2 \hat{g}_{\mu\nu} \nonumber \\
C_{\mu\nu} & = & \hat{\nabla}^2 \left( \hat{R}_{\mu\nu} - \frac{1}{2}  \hat{R} \, \hat{g}_{\mu\nu} \right) - 4 \hat{R}_{\mu\alpha} \hat{R}^{\alpha}_{~~\nu} + 2 \hat{R} \hat{R}_{\mu\nu}  + \frac{3}{2} \hat{R}_{\alpha\beta} \hat{R}^{\alpha\beta} \hat{g}_{\mu\nu} 
- \frac{3}{4} \hat{R}^2 \hat{g}_{\mu\nu}  
\end{eqnarray}
possible in 3-dimensions which come from variation of $\int d^3x\sqrt{\hat{g}}\hat{R}^2$ and $\int d^3x\sqrt{\hat{g}}\left( \hat{R}_{\mu\nu}\hat{R}^{\mu\nu}-\frac{1}{2}\hat{R}^2\right)$ respectively. The coefficients in this expansion, ie. $\rho, p, a, b, \ldots$, are functions only of the (constant) temperature $T$.\footnote{It is perhaps simpler to consider the renormalised one-point function although as stressed earlier, for our application we may equally consider the unrenormalised one since UV divergences will cancel in the quantity of interest, $\Delta F$. In this case one may regard the coefficients $\rho$, $p$ and $a$ as divergent with the UV cut-off.}
Here `$\ldots$' represents higher derivative contributions, terms involving derivatives of $u^\mu$ or temperature, and terms that vanish for our ultrastatic geometry. Note we have used the fact that the Riemann tensor is determined by the Ricci tensor in 3-dimensions, and hence doesn't appear. Also we have no terms linear in $u_\mu$ as we require a static stress tensor, and for our ultrastatic geometry $u_\mu u_\nu \hat{R}^{\mu\nu}$ vanishes, as does $u^\alpha \hat{\nabla}_\alpha \hat{R}$ and $u^\alpha \hat{\nabla}_\alpha \hat{R}_{\mu\nu}$.

Now to compute $\Delta F$ taking $\bar{\Sigma}$ as flat space, and ${\Sigma}$ a weakly curved deformation of it, we may use equation~\eqref{eq:varyDF} and hence only require the spatial components of the stress tensor. These simplify considerably when we write them covariantly in the 2-d geometry $\Sigma$. The terms multiplying $u_\mu u_\nu$ now play no role, and we recall in 2-dimensions the Ricci tensor is given in terms of the Ricci scalar. We obtain the hydrostatic spatial stress tensor components,
\begin{eqnarray}
\langle T_{ij} \rangle_\Sigma &=&  p \, {g}_{ij} + b \left( {\nabla}_i \partial_j {R} - {\nabla}^2 {R} {g}_{ij} -  \frac{1}{4} {R}^2 {g}_{ij}  \right)   + \ldots
\end{eqnarray}
where we note that there are no two-derivative contributions to these spatial components, and also only the 4 derivative tensor $B_{\mu\nu}$ contributes.  The dots again refer to terms with higher than four derivatives of the metric, or terms vanishing in the hydrostatic case, so the derivatives of temperature and terms involving the spatial fluid velocity. Then using~\eqref{eq:varyDF} 
and subtracting a suitable `stretching' of the flat reference space so that $\frac{d}{d\epsilon} \mathrm{Vol}(\bar{g}, \epsilon) = \frac{d}{d\epsilon} \mathrm{Vol}(g, \epsilon)$, we obtain,
\begin{eqnarray}
\frac{d \Delta F}{d\epsilon} &=&   \frac{b}{2}  \int d^2x \sqrt{g} \left( {\nabla}_i \partial_j {R} - {\nabla}^2 {R} \, {g}_{ij} -  \frac{1}{4} {R}^2 {g}_{ij}  \right) \frac{d g^{ij}}{d\epsilon} 
\end{eqnarray}
where we note that the pressure term drops out. This can be integrated in $\epsilon$ to obtain $\Delta F$, and since curvature vanishes for the flat reference space $\bar{\Sigma}$, we obtain, 
\begin{eqnarray}
\label{eq:hydrolimit}
\Delta F[\Sigma] =   \left. - \frac{b(T)}{4}  \int d^2x \sqrt{g} {R}^2 \right|_\Sigma
\end{eqnarray}
for the free energy difference relative to flat space. If we take $g_{ij}(x) = \left( 1 + 2 \epsilon f^{(1)}(x) + O(\epsilon^2) \right) \delta_{ij}$ as above, then the leading order variation yields,
\begin{eqnarray}
\Delta F[\Sigma] =   - \epsilon^2 b(T)  \int d^2x \, (\partial^2 f^{(1)})^2 
\end{eqnarray}
and hence the function $a(k, T)$ in equation~\eqref{eq:quadvary} is determined by the coefficient $b(T)$ as,
\begin{eqnarray}
a(k, T) = 4 \pi^2 b(T) k^4 \, .
\end{eqnarray}
By comparison with the long wavelength results above in equations~\eqref{eq:longwaveholo}, ~\eqref{eq:longwavescalar} and~\eqref{eq:longwavefermion} we can read off the coefficient $b(T)$ for these theories, which is always positive. The expression for $\Delta F$ in equation~\eqref{eq:hydrolimit} is manifestly negative in these cases where the coefficient $b(T)$ is positive. Hence for all these theories we arrive at the interesting result that the perturbative results on the negativity of $\Delta F$ extend to non-perturbative deformations of flat space, provided the hydrostatic limit holds so that the curvature length scale of the deformation is long compared to the thermal length scale. 

We have inferred that the free scalar and fermion QFTs have a hydrostatic description of the free energy difference from the behaviour $a(k, T) \sim k^4$. However, one might ask whether this can be derived directly. We shall now show this is indeed the case. The partition function is given in terms of a functional determinant,
\begin{eqnarray}
 F[\Sigma] = - \tilde{q} \, T \ln \det \hat{\mathcal{O}}_\mathrm{s,f}
\end{eqnarray}
where~$\tilde{q} = -\frac{1}{2}, + \frac{1}{2}$ in the scalar and fermion cases respectively and~$\hat{\mathcal{O}}_\mathrm{s,f}$ are the (2+1)-dimensional scalar and fermion operators,
\begin{eqnarray}
\hat{\mathcal{O}}_\mathrm{s} =- \hat{\nabla}^2 + \xi \hat{R} + M^2 \; , \quad  \hat{\mathcal{O}}_\mathrm{f} = - \hat{\slashed{D}}^2 + M^2 \, .
\end{eqnarray}
In our Euclidean ultrastatic setting one finds,
\begin{eqnarray}
\hat{\mathcal{O}}_\mathrm{s,f} = -\partial_\tau^2 + M^2 +\mathcal{O}_{\mathrm{s,f}}
\end{eqnarray}
where now $\mathcal{O}_\mathrm{s,f}$ are the elliptic 2-dimensional operators on $\Sigma$ given as,
\begin{eqnarray}
\mathcal{O}_\mathrm{s} =- \nabla^2 + \xi R \; , \quad \mathcal{O}_\mathrm{f} = -\slashed{D}^2
\end{eqnarray}
and $\slashed{D}$ is the 2-dimensional Dirac operator on $\Sigma$. The functional determinant may be evaluated via heat kernels in a similar manner to that in~\cite{FisWalWis18}, yielding,
\begin{eqnarray}
\label{eq:DeltaFofell}
\Delta F[\Sigma] = \tilde{q} \, T \int_0^\infty \frac{dt}{t} e^{-M^2 t } \Theta_q(T^2 t ) \Delta K_\mathcal{O_\mathrm{s,f}}(t)
\end{eqnarray}
where~$ \Delta K_\mathcal{O}(t)= \mathrm{Tr}\left(e^{-t\mathcal{O}}-e^{-t\bar{\mathcal{O}}}\right)$ is the difference between the heat kernels for~$\mathcal{O}$ on~${\Sigma}$ and~$\bar{\Sigma}$. 
These  admit an asymptotic heat kernel expansion~\cite{Vas03},
\begin{eqnarray}
\label{eq:HeatKerExp}
K_{\mathcal{O}}(t) \simeq \sum_{m\geqslant 0 } d_{2m}(\mathcal{O} )t^{m-1}
\end{eqnarray}
where~$d_{2m}(\mathcal{O})$ are the heat kernel coefficients, and in our 2-dimensional context these are integrals of sums and products of the Ricci scalar and its derivatives so that~$d_{2m}(\mathcal{O}) \sim \ell^{2-2m}$, where~$\ell$ is the characteristic length scale of the deformation. Hence this should be viewed as an expansion in the dimensionless quantity $t/\ell^2$.
In the hydrostatic regime discussed above we have low curvature compared to the thermal scale, so that~$\ell \, T \gg 1$. Then the integrand in equation~\eqref{eq:DeltaFofell} is localised near~$t=0$ (where in the scalar case we also require a non-zero mass such that~$\ell \, M \gg 1$) and using the heat kernel expansion gives,
 \begin{eqnarray}
\Delta F \simeq \frac{\tilde{q}}{2} \sum_{m=0}^{\infty}  \Delta d_{2m+4}^{\mathrm{s,f}} \frac{(-1)^m }{T^{2m+1}} J_{\mathrm{s,f}}^{(m)}\left( \frac{M^2}{T^2} \right) 
\end{eqnarray}
where,
 \begin{eqnarray}
J_{\mathrm{s}}(x) = \frac{1}{\sqrt{x}} \coth\left( \frac{\sqrt{x}}{2} \right) \; ,\quad J_{\mathrm{f}}(x) = \frac{1}{\sqrt{x}} \tanh\left( \frac{\sqrt{x}}{2} \right) 
\end{eqnarray}
and we have used the fact that the first two heat kernel coefficients are proportional to the volume and Euler characteristic of~$\Sigma$, respectively, and cancel in the free energy difference when computed using the `stretched' reference flat space metric, $\bar{g}_{ij}(x ; \epsilon)$, as detailed in section~\ref{sec:variation}.
The first term in the expansion is then determined by the heat kernel coefficients~\cite{Vas03},
\begin{eqnarray}
d_{4}^\mathrm{s} = \frac{1}{240\pi}\left(1-10\xi+30\xi^2\right)\left.\int d^2 x \sqrt{g} R^2\right|_{\Sigma} \; , \quad
d_{4}^\mathrm{f} = -\frac{1}{480\pi}\left.\int d^2 x \sqrt{g} R^2\right|_{\Sigma} 
\end{eqnarray}
for the 2-dimensional scalar and Dirac operators, respectively. These give a leading contribution to $\Delta F$ that agrees precisely with the expressions obtained by using the form derived from hydrostatics in equation~\eqref{eq:hydrolimit}, and the value of the coefficient $b(T)$ determined from the perturbative results in equations~\eqref{eq:longwavescalar} and~\eqref{eq:longwavefermion}.

\section{Summary and discussion}

We have defined the free energy difference $\Delta F(\epsilon) = F(\epsilon) - \bar{F}(\epsilon)$ where the free energy of a relativistic QFT on the product of time with a space $\Sigma(\epsilon)$ with metric $g_{ij}(x; \epsilon) = e^{2 f(x ; \epsilon)} \bar{g}_{ij}(x)$ is subtracted from that on a reference geometry $\bar{\Sigma}(\epsilon)$ at the same temperature. If the geometry is compact we require $\bar{\Sigma}$ to have metric $\bar{g}_{ij}(x)$ and choose the deformation $f(x ; \epsilon)$ so that the volumes $\mathrm{Vol}(g)$, $\mathrm{Vol}(\bar{g})$ are equal. In the non-compact case where $\Sigma(\epsilon)$ is a deformation of flat space, we subtract an appropriately `stretched' reference flat space with spatial metric $\bar{g}_{ij}(x ; \epsilon)$ (in the same coordinates as for $g_{ij}$) given by an $\epsilon$ dependent diffeomorphism of the flat reference metric $\bar{g}_{ij}(x)$, with the diffeomorphism chosen so that the volume variations $\frac{d}{d\epsilon} \mathrm{Vol}(g, \epsilon)$ and $\frac{d}{d\epsilon} \mathrm{Vol}(\bar{g}, \epsilon)$ are equal. For a (2+1)-dimensional QFT (although not for higher dimensions) this quantity $\Delta F(\epsilon)$ is UV finite, and hence is independent of renormalization 
ambiguity.
We expect that for reasonable fall off of the perturbation this will also be IR finite on non-compact spaces.

We have shown that for deformations which are perturbations of flat space, then under the assumption that for flat space the finite temperature spatial stress tensor is governed by a homgoeneous isotropic pressure, then the variation of $\Delta F(\epsilon)$ is quadratic in the perturbation.
We have also shown that the same holds for perturbations of homogeneous compact spaces, where for the unperturbed homogeneous space the trace of the spatial parts of the stress tensor, $\bar{\sigma}^i_{~i}$, is constant -- as for example we would expect on a sphere or torus. This explains why the variation of free energy found on a torus in~\cite{FisWalWis18} was quadratic.

In the compact case the free energy difference $\Delta F(\epsilon)$ is a physically relevant quantity. Imagine we could design a compact 2-dimensional membrane in the lab on which degrees of freedom live that are governed by a relativistic QFT on the curved ultrastatic geometry induced on the membrane. As discussed in~\cite{FisWalWis18}, monolayer graphene is precisely thought to be such a material. 
To be more concrete let us consider deformations of a spherical membrane. Take flat Euclidean 3-space in spherical coordinates $ds_{(3)}^2 = dr^2 + r^2 d\Omega^2$ and consider the round embedding $r = R$, giving the induced geometry $\Sigma(0)$ with metric $ds^2 = R^2 (d\theta^2 + \sin^2{\theta} d \phi^2) = \bar{g}_{ij} dx^i dx^j$, and we take $x^i = (\rho,\phi)$ with $\rho = R \theta$. Note that near the pole, $\rho = 0$, then $ds^2 \simeq d\rho^2 + \rho^2 d\phi^2$ and we use this shortly.
Now perturb this embedding as $r = R ( 1 + \epsilon f^{(1)}(\rho,\phi) )$ so that the sphere area is preserved. The induced geometry $\Sigma(\epsilon)$ has metric,
\begin{eqnarray}
ds^2 = g_{ij}( x ; \epsilon ) dx^i dx^j = ( 1 + 2 \epsilon f^{(1)}(x) ) \bar{g}_{ij} dx^i dx^j + O(\epsilon^2)
\end{eqnarray}
precisely of the form we have considered, and area preservation implies $\int d^2x \sqrt{\bar{g}} f^{(1)}(x) = 0$.
If the membrane were inextensible it would necessarily be that deformations preserve area. 
Let us suppose that the system may be thought of in a Born-Oppenheimer approximation, as being well described by  `heavy'  non-relativistic degrees of freedom of the membrane together with  `light' relativistic QFT degrees of freedom. Then we may `integrate out' (or solve) the QFT to derive a free energy as a functional of the fixed `heavy' degrees of freedom, the membrane geometry. 
Then we may think of this QFT free energy as giving a potential for the `heavy' membrane degrees of freedom. 
Suppose at a given temperature the spherical membrane is an equilibrium configuration. Then our quantity $\Delta F(\epsilon)$ precisely describes the resulting quadratic `potential' free energy contribution as the membrane is perturbed, provided it is inextensible, or is deformed to preserve its area. 
The membrane may have other potential contributions, for example a bending energy. One would then have to consider whether $\Delta F(\epsilon)$ is an important contribution. This was discussed in~\cite{FisWalWis18} will be further considered in~\cite{ongoing}.

Here we have focused on computing the quadratic variation of $\Delta F$ not in the compact case, but rather the non-compact case for deformations of flat space. We have computed this variation for holographic CFTs, and shown that the quadratic variation is always negative for any temperature and perturbation, as was previously found for free scalar and fermions in~\cite{FisWalWis18}. Furthermore it is strikingly similar in functional form to that for the free massless Dirac fermion CFT (the theory on the worldvolume of graphene). We have also argued that in any theory that has a hydrostatic limit for long wavelength deformations of flat space (but now not necessarily small in amplitude) relative to the temperature scale, then $\Delta F$ is governed by a specific curvature correction to the stress tensor, and is negative for the free scalar and fermion theories, as well as the holographic CFT.

For  perturbations of flat space we have that $\Delta F$ takes the form,
\begin{eqnarray}
\Delta F(\epsilon) & = & - \epsilon^2 \int d^2k \, a( k ) \left| \tilde{f}(k_i) \right|^2 + O\left( \epsilon^3 \right)
\end{eqnarray}
and $a(k)$ for the holographic CFT computed here, and for free fields computed previously, decays as $a(k) \sim k^4$ for long wavelengths. This implies that the quadratic variation is dominated by shorter wavelength deformations. 
With this in mind, more generally for perturbations of a compact space, we expect that the quadratic variation of free energy will behave as our flat space quantity $\Delta F$ for perturbations of the geometry that are on scales less than the curvature scale of the space. Long wavelength parts of the perturbation may result in the perturbation preserving area, leading to a quadratic variation of free energy for a homogeneous space such as a sphere, but the short distance variation of free energy will be approximately that of our quantity $\Delta F$ for flat space. 
Let us return to our spherical membrane example.
Consider a perturbation $f^{(1)}$ which is inhomogeneous only in a small region of the sphere, $\rho < L \ll R$, and is constant outside this region.
We define $\alpha = L/R$ which we take to be small.\footnote{We should be careful that whilst the ratio of $L$ to $R$ is small, we still require $\alpha = \frac{L}{R} \sim O(1)$ as $\epsilon \to 0$ otherwise the second order perturbation to the metric may become larger than $O(\epsilon^2)$ due to derivatives of $f^{(1)}$ in those terms.} For example we might take the axially symmetric perturbation,
\begin{eqnarray}
f^{(1)}(x^i) = \left\{ \begin{array}{cc}
c + e^{-\tan^2\left( \frac{\pi \rho}{2 L} \right)} & \rho < L \\
c & \rho \ge L \\
\end{array} \right.
\end{eqnarray}
Then area preservation implies the constant $c$ is non-zero and negative, and parametrically goes as $c \sim - \alpha^2$. We expect the response of free energy to be dominated by the short distance variation of this perturbation, and the large scale constant part given by $c$ that results in volume preservation will be irrelevant. Thus in this case the free energy variation as the sphere is deformed would be approximately given by the flat space free energy variation $\Delta F$ for a flat space perturbation $f^{(1)}(x) = e^{-\tan^2\left( \frac{\pi \rho}{2 L} \right)}$ in polar coordinates $ds^2 = d\rho^2 + \rho^2 d\phi^2$ (which we note does not preserve area).

Thus we may ultimately view our formal flat space quantity $\Delta F$ as giving the short distance response of thermal vacuum energy on scales shorter than the characteristic curvature scale of the space that is perturbed.\footnote{Note that for temperature to be relevant in this limit, it must also be large compared to the curvature scale of the unperturbed space.} The fact that in the theories discussed -- the free fields and holographic CFT -- the contribution is always negative for perturbations of flat space is then a statement more generally about perturbations on short scales on \emph{any} space giving rise to negative free energy variations. In the physical context of membranes discussed above this may have physical implications for their dynamics and lead to a tendency to crumple, as discussed in~\cite{FisWalWis18}, if other contributions to the membrane dynamics such as bending energy do not counteract this.



\section*{Acknowledgements}

We would like to thank Sebastian Fischetti for many valuable conversations. TW would like to thank Shinji Mukohyama and the YITP for their kind hospitality while this work was completed. This work was supported by the STFC grant ST/P000762/1 and an STFC studentship. KC is sponsored by the DPST scholarship from the Royal Thai Government.


\bibliographystyle{apsrev4-1}
\bibliography{spectrum}

\end{document}